\documentclass[aps,prl,reprint,showpacs,twocolumn,superscriptaddress]{revtex4-1}

\usepackage[dvips]{graphicx}
\usepackage{dcolumn}
\usepackage{bm}
\usepackage{xspace}
\usepackage{multirow}
\usepackage{natbib}
\usepackage{color}

\begin{document}

\preprint{}
\title{Disappearance of Superconductivity Due to Vanishing Coupling in the Overdoped Bi$_2$Sr$_2$CaCu$_2$O$_{8+\delta}$}
\author{T. Valla}
\author{I. K. Drozdov}
\author{G. D. Gu}
\affiliation{Condensed Matter Physics and Materials Science Department, Brookhaven National Lab, Upton, New York 11973, USA\\}

\date{\today}

\begin{abstract}
In high-temperature cuprate superconductors, superconductivity is accompanied by a \lq{}plethora of orders\rq{}, and phenomena that may compete, or cooperate with superconductivity, but which certainly complicate our understanding of origins of superconductivity in these materials. While prominent  in the underdoped regime, these orders are known to significantly weaken or completely vanish with overdoping. Here, we approach the superconducting phase from the more conventional highly overdoped side. We present angle-resolved photoemission spectroscopy (ARPES) studies of Bi$_2$Sr$_2$CaCu$_2$O$_{8+\delta}$ (Bi2212) single crystals cleaved and annealed in ozone to increase the doping all the way to the metallic, non-superconducting phase. We show that the mass renormalization in the antinodal region of the Fermi surface, associated with the structure in the quasiparticle self-energy, that possibly reflects the pairing interaction, monotonically weakens with increasing doping and completely disappears precisely where  superconductivity disappears. This is the direct evidence that in the overdoped regime, superconductivity is determined by the coupling strength. A strong doping dependence and an abrupt disappearance above the transition temperature ($T_{\mathrm c}$) eliminate the conventional phononic mechanism of the observed mass renormalization and identify the onset of spin-fluctuations as its likely origin.
\end{abstract}
\vspace{1.0cm}


\maketitle\section*{introduction}
More than 30 years after the discovery of cuprate superconductors, the pairing mechanism in these materials still remains unknown. The observation of renormalization effects in the low energy electronic excitations in ARPES has re-ignited the hope that a bosonic mode playing a role in pairing in cuprates could finally be identified, in analogy with how tunneling experiments provided the smoking gun evidence for phononic mechanism in conventional superconductors \cite{McMillan1965}. However, after two decades of intense research, the debate about the coupling mechanism is still open \cite{Valla1999,Johnson2001a,Gromko2003,Carbotte2011,Park2013,Li2018,He2018}. One problem was that early studies were focused on the nodal "kink" that did not show any significant correlations with superconductivity when the latter was altered by doping or when different cuprate families were compared. Another problem is that cuprates are fundamentally different from simple metals in which superconducting transition occurs from a conventional Fermi liquid metallic state into a state well described by the BCS theory \cite{Bardeen1957a,Bardeen1957}. Parent compounds of cuprate superconductors are antiferromagnetically ordered Mott insulators wherein conduction and superconductivity are induced by doping additional holes or electrons away from the half filled case \cite{Lee2006}. The effects of strong correlations extend far away from half filling, deep into the regime that overlaps with superconductivity, where their presence and intertwining with superconductivity complicates the identification of the superconducting mechanism. Therefore, it would be desirable to study superconducting properties in the highly overdoped regime where such effects are absent or strongly reduced.

Bi2212 has been a perfect subject of ARPES studies due to its ease of cleaving, a high transition temperature ($T_\mathrm{c}$), and a large superconducting gap. However, Bi2212 could only be doped within a relatively limited range on the overdoped side, where $T_c$ could not be reduced below $\sim50$ K, leaving a crucially important region of the phase diagram, where $T_\mathrm{c}\rightarrow{0}$, out of reach. Only very recently, has it become possible to extend the overdoped range beyond the point at which superconductivity vanishes by annealing the \textit{in-situ} cleaved samples in ozone \cite{Drozdov2018}. For the first time, this has made it possible to monitor the development of electronic excitations as superconductivity weakens and finally completely disappears, allowing a closer look at its origins. 

\section*{Results}

\begin{figure}[htpb]
\begin{center}
\includegraphics[width=9cm]{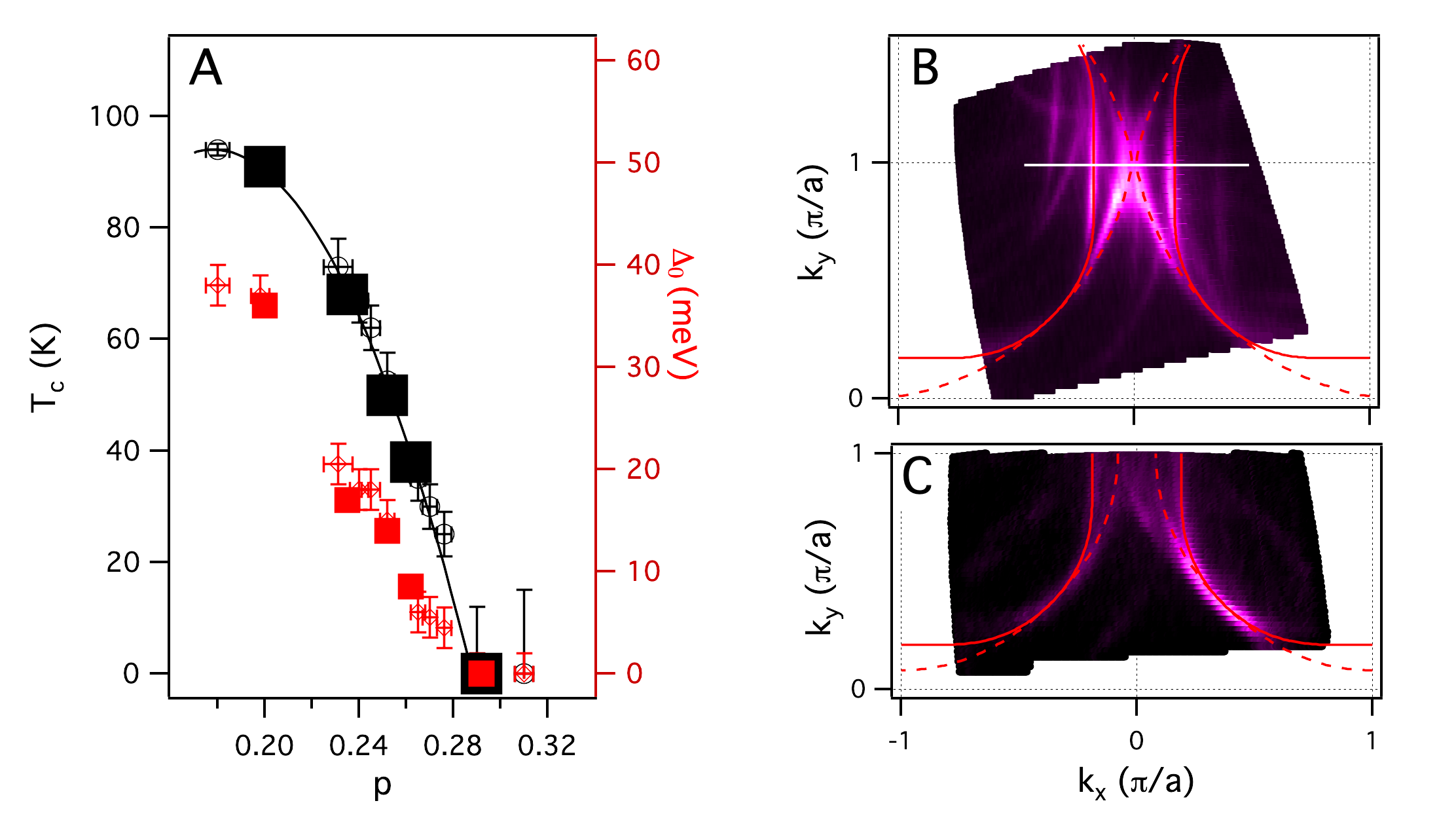}
\caption{Strongly overdoped regime of Bi$_2$Sr$_2$CaCu$_2$O$_{8+\delta}$. (A) Phase diagram near the edge of the superconducting dome, as determined from ref. \cite{Drozdov2018}. $T_\mathrm{c}$ and $\Delta_0$ for the doping levels from this study are indicated by the black and red solid squares, respectively. (B) Fermi surface ($E=0$ contour) of the overdoped, non-superconducting sample, corresponding to $p=0.29$ and (C) of the $T_\mathrm{c}=72$ K sample, corresponding to $p=0.23$. Maps in (B) and (C) were recorded at $T=12$ K. 
}
\label{Fig1}
\end{center}
\end{figure}

Figure 1(A) shows the overdoped region of the Bi2212 phase diagram from ref. \cite{Drozdov2018}, along with the four doping levels from the present study. 
In this region, the pseudogap is no more present, according to the previously published studies \cite{Renner1998,Ozyuzer2000,Yusof2002,Gomes2007,Benhabib2015} and the remaining  superconductivity becomes more conventional with the gap saturating at the BCS value $2\Delta_0=4.28k_BT_c$ for $p>0.25$ \cite{Drozdov2018,He2018}. 
The as grown OD91 $(p=0.2)$ sample was cleaved in vacuum and annealed in ozone, resulting in increased doping, $p=0.29$, and a complete loss of superconductivity. The Fermi surface of the resulting sample is shown in Fig. 1(B). That same sample is then annealed in vacuum at different temperatures, ranging from 110 to 175$^{\circ}$ C in order to gradually reduce the doping and increase $T_\mathrm{c}$ to 38, 50 and 72 K. The intensity at the Fermi level of the same surface after the final annealing is shown in Fig. 1(C). Due to the large superconducting gap ($\Delta_0=17$ meV), the photoemission intensity is concentrated near the nodes. The doping level in each case is determined independently from the Luttinger count of the area enclosed  by the Fermi contour, $p_\mathrm{L}=2A_\mathrm{FS}$. The doping $p$ that serves as the abscissa in phase diagrams of the cuprates, (the doping away from the half-filling) is expressed as $p=p_\mathrm{L}-1=2A_\mathrm{FS}-1$ with both the bonding and the antibonding states counted, $A_\mathrm{FS}=(A_\mathrm{B}+A_\mathrm{A})/2$. The area of the Brillouin zone (BZ) is set to one. 
Also shown are the Fermi surface contours of the tight-binding (TB) in-plane band structure that best describe the measured ones, as described in the Methods section. 

The antinodal gap magnitude $\Delta_0$ is determined at the base temperature $(T\approx 12$ K) from the quasiparticle peak position at $(0, \pm k_F)$, while the transition temperature $T_\mathrm{c}$ is determined as the temperature at which the gap closes. The points from the present study shown in Fig. 1(A) follow the trends from our previous study \cite{Drozdov2018}. Indeed, for the initial ozone annealed surface, that shows no superconductivity within our detection limits, the Van Hove singularity of the antibonding state sits exactly at the Fermi level. 

\begin{figure*}[htpb]
\begin{center}
\includegraphics[width=12cm]{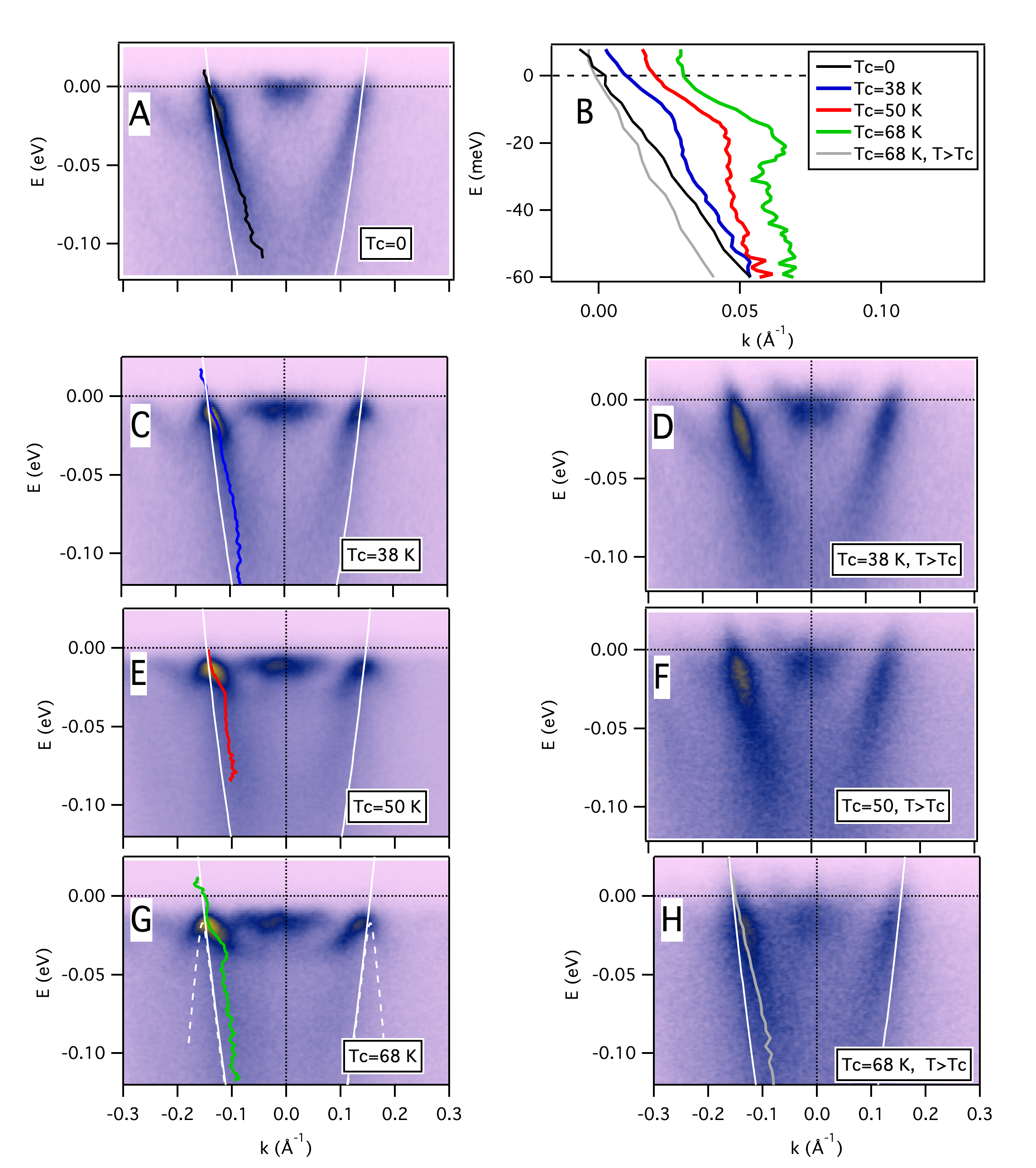}
\caption{Coupling strength in the overdoped Bi2212 as a function of doping. (A) Electronic structure of Bi2212 near the antinode along the momentum line indicated in Fig.\ref{Fig1}(B) at low temperature ($T\sim10$ K) for overdoped, non-superconducting sample. The spectra corresponding to the three overdoped superconducting samples with $T_\mathrm{c}=38$ K, $T_\mathrm{c}=50$ K and $T_\mathrm{c}=72$ K taken in the superconducting state (C, E, G) and normal state (D, F, H). The MDC-fitted dispersions of the bonding state are indicated by the black, blue, red, green and gray curves. The TB dispersions are indicated by the solid white curves. The dashed white curve in (G) represents the TB dispersion gapped by $\Delta_0=17$ meV. (B) The same measured dispersions, referenced to the corresponding gap value. The momentum scale is referenced to $k_\mathrm{F}$. The dispersions corresponding to superconducting states are offset in $k$ by 0.01 \AA, consecutively. Spectra in (A), (C), (E) and (G) were recorded at $T=12$ K and those in (D), (F) and (H) at 45, 60 and 90 K, respectively.
}
\label{Fig2}
\end{center}
\end{figure*}

This is also illustrated in Fig. 2(A) that shows the photoemission intensity along the momentum line $k_\mathrm{y}=\pi/a$ indicated by the yellow line in Fig. 1(B). The state at $(0,\pi/a)$ is the bottom of the antibonding band that undergoes a Lifshitz transition at that doping level $(p=0.29)$. The remaining state, that crosses the Fermi level at $k_\mathrm{F}=\pm 0.144$ \AA\  is the bonding state. Its dispersion (black curve), extracted by fitting the momentum distribution curves (MDC), does not show any features that would indicate a structure in the self energy and a renormalization in the form of a "kink".  Still, the dispersion is slightly renormalized compared to the TB approximation that was used for the Fermi surface contour (Fig. 1(B)). The state is gapless and does not show any particle-hole mixing expected for Bogoliubov's quasiparticles in the superconducting state. With vacuum annealing and a reduction in hole doping, superconductivity develops and the spectra display the spectral gap at low-temperatures (panels C, E and G). Simultaneously, the photoemission shows a back-folding of the spectral intensity near the $k_\mathrm{F}$, typical for Bogoliubov's quasiparticles. However, the most important discovery here is an anomaly, or an abrupt change of slope ("kink") in the state's dispersion that occurs slightly below the state's maximum at $k_\mathrm{F}$. This can be seen in the MDC-derived dispersions, represented by blue, red and green curves for the samples with $T_\mathrm{c}$ of 38, 50 and 72 K, respectively. When plotted on the same scale and referenced to the corresponding gap magnitude, panel (B), these dispersions indicate clear trends in their low-energy behavior: as superconductivity strengthens and $T_\mathrm{c}$ and $\Delta_0$ increase, the "kink" becomes progressively more pronounced and shifts to higher energies. Notably, the "kink" is present only in the superconducting state with no traces of the structure left above $T_\mathrm{c}$, as can be seen in the corresponding normal state spectra taken approximately 10 K above $T_\mathrm{c}$ (panels D, F and H). 
This is highly unusual and, as already noted in previous studies \cite{Valla2000,Gromko2003,Li2018}, cannot be reconciled with the conventional effects stemming from the electron-phonon coupling. If the "kink" was due to the conventional electron-phonon coupling that is at play in 2H-NbSe$_2$ and intercalated graphite, for example, it would have to be present not only in the superconducting state, but also should exist in the normal state \cite{Valla2004,Valla2009}, as illustrated in Fig. 3(E).  
 
To quantify the trends observed in Fig. 2, we plot the kink's characteristic energy, $\Omega_0$, corresponding to the maximum in Re$\Sigma$ and its strength, approximated by $\lambda=-\frac{\partial\mathrm{Re}\Sigma(\omega)} {\partial\omega} \vline_{(\Omega_0<\omega<\Delta_0)}$, in Figure 3. In addition, we re-plot the corresponding maximal gap, $\Delta_0$, and show the energy of the resonance mode, $E_\mathrm{r}$, and a spin gap, $\Delta_\mathrm{Spin}$, from the inelastic neutron scattering studies \cite{Fong1999,He2001,Capogna2007,Fauque2007,Xu2009,Li2018a}. The energy of the B$_\mathrm{1g}$ phonon is also indicated, noting that it does not show a significant doping dependence \cite{Opel1999}. 
We note that a weak featureless renormalization remains at $p=0.29$ and in the normal state of superconducting samples. That component does not display any doping dependence in the studied range. We call the corresponding slope of Re$\Sigma$ the critical coupling, $\lambda_\mathrm{c}$ as the $p=0.29$ sample sits exactly at the superconducting boundary. 

\begin{figure*}[htbp]
\begin{center}
\includegraphics[width=15cm]{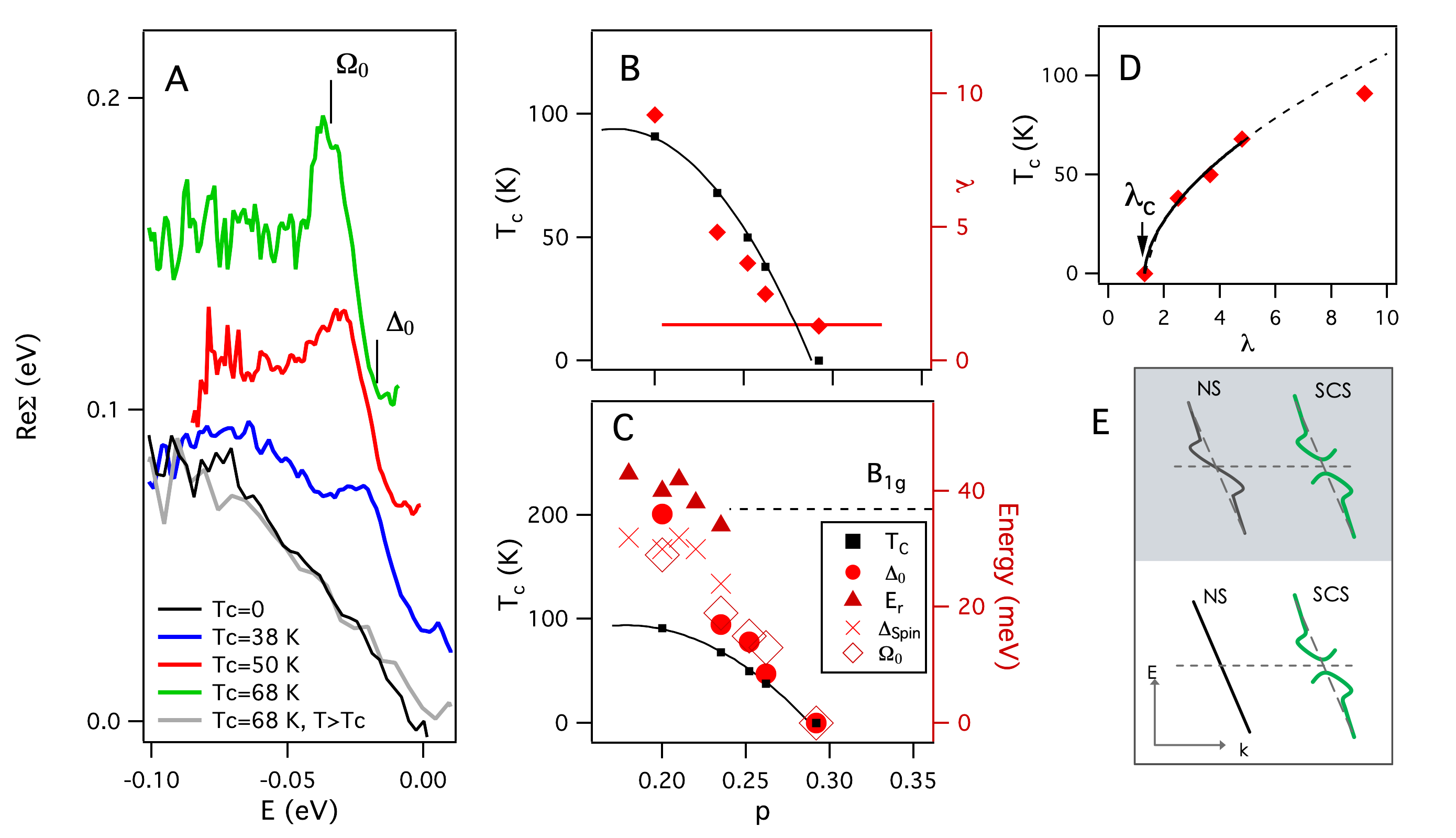}
\caption{Doping dependence of the antinodal renormalization effects. (A) Re$\Sigma$ for four samples shown in Fig.\ref{Fig2} obtained by subtracting the bare TB dispersion, gapped by the corresponding $\Delta_0$, from each measured dispersion. The curves are referenced to the Fermi level and those obtained in superconducting state are offset in $y$ by 30 meV for clarity. (B) coupling strength $\lambda$, approximated as $\lambda=-\frac{\partial\mathrm{Re}\Sigma(\omega)} {\partial\omega}\vline_{(\Omega_0<\omega<\Delta_0)}$ (red diamonds), plotted vs. doping. The normal state  value, $\lambda_\mathrm{c}\approx1.3$, is indicated by the red line. Corresponding $T_\mathrm{c}$ is also shown (black squares). (C) Kink's energy, $\Omega_0$, as measured from the corresponding gap value (energy of the maximum in the state's dispersion) (red diamonds). Corresponding gap magnitude, $\Delta_0$ (red circles) of the studied samples and antiferromagnetic resonance energy, $E_\mathrm{r}$ (red triangles), and spin gap, $\Delta_\mathrm{Spin}$ (red crosses), from references \cite{Fong1999,He2001,Capogna2007,Fauque2007,Xu2009,Li2018a} are also shown. (D) Dependence of $T_\mathrm{c}$ on the antinodal coupling strength, $\lambda$, measured in the superconducting state. The solid curve represents the fit to the power-law behavior, $T_\mathrm{c}\propto(\lambda-\lambda_\mathrm{c})^p$ for the four overdoped samples. The dashed curve is the extrapolation from the fitted region. The as grown sample ($T_\mathrm{c}=91$ K) was not used in fitting. (E) Schematic view of temperature development of the electronic dispersion upon transition from the normal state (NS) to superconducting state (SCS) in the conventional coupling scenario (top, shaded) and the actual one, observed in cuprate superconductors (bottom).
}
\label{Fig3}
\end{center}
\end{figure*}

\section*{Discussion}
It is obvious that both the strength of the anomaly and its energy are strongly doping dependent, both following $T_\mathrm{c}$ and and vanishing exactly when superconductivity disappears. This represents very strong evidence that the antinodal kink is very closely related to superconductivity. The fact that $\Delta_0$ and the observed coupling follow $T_\mathrm{c}$ and essentially vanish together at the overdoped side is a clear indication that the superconductivity itself turns conventional in that region of the Bi2212 phase diagram and that it is governed by the weakening coupling, rather than by the superfluid density, as suggested by ref. \cite{Bozovic2016}. 

The antinodal dispersion anomaly also occurs in the $k$-space region where the superconducting gap and pairing are the strongest \cite{Gromko2003,Li2018}. The fact that it only exists in the superconducting state also provides additional clues for understanding its origin. In that, the antinodal kink is strikingly different from the nodal kink, which does not vary significantly with doping or amongst different cuprate families \cite{Lanzara2001,Johnson2001a,Ino2013,Park2013}. The apparent lack of correlation of the nodal kink with $T_\mathrm{c}$ suggests its relative unimportance in superconductivity. The nodal kink is also different in that it exists in both the normal and superconducting states, with only a relatively small change upon the transition,  allowing the possibility that it might be phonon related. In contrast, the strong doping dependence and the simultaneous disappearance of the antinodal kink with superconductivity would require that strong changes in the coupling and in the phonon spectrum itself occur with doping and temperature, if the kink had phononic origin. This has not been observed \cite{Opel1999}.

The recent study on the same material reports that the coupling strength has a similar trend with doping \cite{He2018}. However, that study assigns the observed effects, i.e. the development of the "peak-dip-hump" structure in the spectra at ($\pi,0$), to the coupling to B$_\mathrm{1g}$ phonon whose energy does not vary with doping ($\omega_0=37$ meV). Also, the study does not address a lack of the coupling above $T_\mathrm{c}$. We note that our results, showing strong doping dependence of $\Omega_0$ and a striking change between the superconducting and normal state spectra rule out the possibility that the involved mode is a phonon. As illustrated in Fig. 3(E), if caused by phonons, kink should be present in both the normal and superconducting states.

The second bosonic candidate that is often considered as the origin of the observed quasiparticle kink is the so called spin resonance \cite{Fong1999,Abanov1999,Eschrig2000,He2001,Sandvik2004,Eremin2005,Capogna2007,Dahm2009}. The energy of that mode, $E_\mathrm{r}$, shows the doping dependence with the same trend as the energy of the kink studied here. Also, its temperature dependence is similar, with both phenomena existing only in the superconducting state. However, as Fig. 3(C) shows, there is a significant mismatch between the energies of the two features. The overlapping point between the neutron scattering and ARPES data, corresponding to the $T_\mathrm{c}\approx70$ K sample, would suggest that the $\Delta_0+\Omega_0$ scale from ARPES is a better match to $E_\mathrm{r}$. However, that clearly would not work near the optimal doping. We also note that the momentum and energy conservation rules would have to place the antinodal kink near the energy of the involved mode (as measured from top of electronic dispersion at $\Delta_0$), particularly if the mode scatters from the antinode to the antinode (small $Q$, or $Q\approx(\pi,\pi)$). This is why a much better candidate for the relevant excitation seems to be the onset of spin fluctuation spectrum, i.e. the spin gap ($\Delta_\mathrm{Spin}$), rather than the resonance mode at $E_\mathrm{r}$ \cite{Li2018a}. The excitations at the spin gap could explain not only the kink's doping, temperature and momentum dependence, but also the differences between the different families of cuprates - most notably those between Bi2212 and La$_{2-x}$Sr$_x$CuO$_4$. These two materials have very similar scales for $E_\mathrm{r}$, but a much smaller $\Delta_\mathrm{Spin}\approx8$ meV in La$_{2-x}$Sr$_x$CuO$_4$ near optimal doping would definitely make the observation of a coherent quasiparticle peak and a kink in its dispersion very difficult, in agreement with ARPES measurements \cite{Yoshida2016}.

At the end, the remarkable correlation between $T_\mathrm{c}$ and coupling strength from Fig. 3(B) could offer an interesting insight into the question if the transition temperature in cuprates might reach a limit when coupling gets very strong. When plotted as a function of $\lambda$, transition temperature displays approximately a square-root behaviour on $(\lambda-\lambda_\mathrm{c})$ in the overdoped regime (Fig. 3(D)). This is a good news and an indication that $T_\mathrm{c}$ in cuprates does not have a natural limit in the coupling strength itself. However, on the underdoped side, there are many phenomena that limit $T_\mathrm{c}$, even when coupling is finite, some of these probably being caused by the strong coupling observed here. The point corresponding to the $T_\mathrm{c}=91$ K sample, laying below the extrapolated curve, indicates that this region might already be affected. 

\section*{Methods}

\subsection{Sample Preparation}
The experiments within this study were done in a new experimental facility that integrates oxide-MBE with ARPES and scanning tunneling spectroscopy (STM) capabilities within the common vacuum system \cite{Kim2018a}. The starting sample was a slightly overdoped ($T_c=91$ K) single-crystal of Bi$_2$Sr$_2$CaCu$_2$O$_{8+\delta}$, synthesized by the traveling floating zone method. It was clamped to the sample holder and cleaved with Kapton tape in the ARPES preparation chamber (base pressure of $3\times10^{-8}$ Pa). The silver-epoxy glue, commonly used for mounting samples, as well as the need for its processing at elevated temperatures, have been completely eliminated, resulting in perfectly flat cleaved surfaces and unaltered doping level. The cleaved sample was then transfered to the MBE chamber (base pressure of $8\times10^{-8}$ Pa) where it was annealed in $3\times10^{-3}$ Pa of cryogenically distilled O$_3$ at 350-480$^{\circ}$C for $\approx$ 1 hour. After the annealing, sample was cooled to room temperature in the ozone atmosphere and transfered to the ARPES chamber (base pressure of $8\times10^{-9}$ Pa). No spectral gap was detected down to the base temperature (12 K) and the doping level determined from the area of the Fermi surface was $p=0.29$. Reduction in doping was achieved by subsequent annealing of the same surface in vacuum to temperatures ranging from 110 to 175$^{\circ}$C, resulting in development of superconductivity with increasing $T_\mathrm{c}$. 

\subsection{ARPES}

The ARPES experiments were carried out on a Scienta SES-R4000 electron spectrometer with the monochromatized HeI (21.22 eV) radiation (VUV-5k). The total instrumental energy resolution was $\sim$ 4 meV. Angular resolution was better than $\sim 0.15^{\circ}$ and $0.3^{\circ}$ along and perpendicular to the slit of the analyzer, respectively. 

The annealing of cleaved surfaces in ozone results in increased doping only in the near-surface region, while the subsequent annealing in vacuum reduces it. Therefore, aside from the as-grown sample, the only measure of $T_c$ in the near-surface region was spectroscopic: the temperature induced changes in the quasiparticle peak intensity, as well as the leading edge position indicate $T_c$ \cite{Damascelli2003,Kondo2015}. The leading edge gap and intensities of the QP peak and at the Fermi level all show a prominent change around $T_c$ and the later could be identified as being near the inflection point of these temperature dependencies \cite{Kondo2015}. The ARPES estimate of $T_\mathrm{c}$ was within $\pm4$ K, except for the sample falling outside of the superconducting dome, for which the estimate was limited by the base temperature that could be reached with our cryostat ($T_\mathrm{c}<12$  K).

\subsection{As-grown sample}

\begin{figure}[htpb]
\begin{center}
\includegraphics[width=8cm]{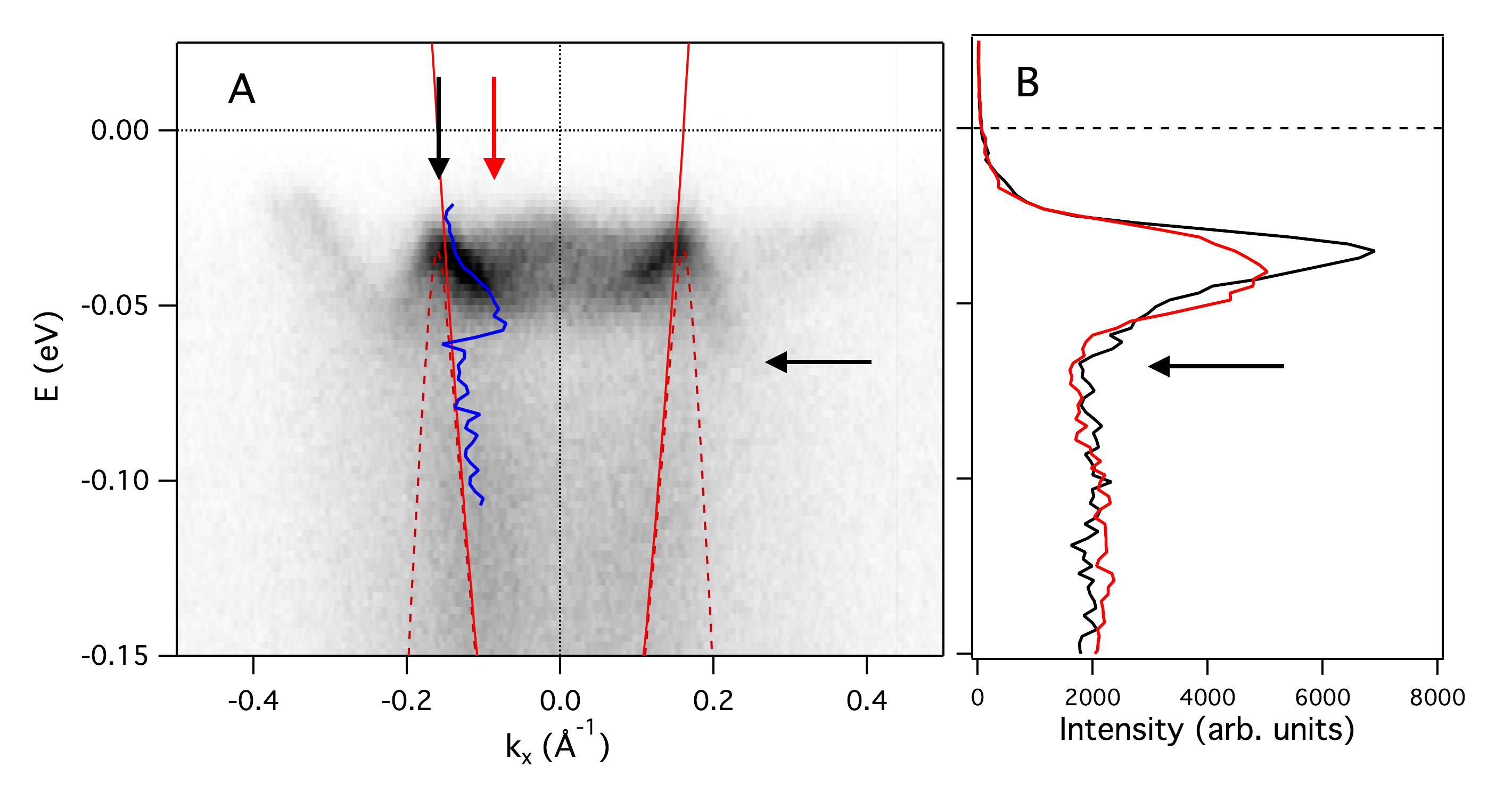}
\caption{As grown Bi2212 sample ($T_\mathrm{c}=91$ K). (A) Electronic structure near the antinode along the momentum line indicated in Fig. 1(B) at low temperature ($T\sim12$ K) for the as-grown Bi2212 sample. The MDC-fitted dispersions of the bonding state is indicated by the blue curve. The TB dispersion is indicated by the solid red curve. The dashed red curve represents the TB dispersion gapped by $\Delta_0=35$ meV. (B) The energy distribution curves corresponding to the $k_\mathrm{F}$ (black) and the momentum indicated by the red vertical arrow in (A). The horizontal black arrow indicates the ``dip'' in the intensity.
}
\label{Fig4}
\end{center}
\end{figure}

The spectra for the as-grown, slightly overdoped ($T_c=91$ K) sample (Fig. 4) cannot be reliably analyzed in the same manner as the spectra for highly overdoped samples. The MDC analysis returns a well defined result for the state's dispersion in the low-energy range and in the high-energy range, but not in the vicinity of the kink. This is partially due to the fact that on the particle-like side ($|k|<k_\mathrm{F}$) of the renormalized Bogoliubov's dispersion, the two sides corresponding to negative and positive momenta, merge and form a continuous renormalized dispersion, with the bottom at $k_x=0$ that could be shallower than the energy of the re-normalizing mode. Also, the intensity from the antibonding state and super-modulation replicas partially overlaps with the fitted state and the MDC fitting is unstable and often shows a sharp discontinuity near the kink energy. Obviously, the energy of the kink cannot be  precisely established by using the MDC analysis, whereas the low-energy slope, that serves for determination of the coupling strength $\lambda$, can still be correctly determined. Therefore, for the lower limit of the mode's energy we use the energy at which the MDC derived dispersion (blue curve in Fig. 4(A)) shows a discontinuity. As its upper limit, we use the energy at which the energy distribution curves show a ``dip'' (Fig. 4(B)).  This energy coincides with the energy within which the hole-like portion($|k|>k_\mathrm{F}$) of the Bogoliubov's dispersion shows the ``heavy'', renormalized character.  That part  of the renormalized Bogoliubov's dispersion could be traced all the way to the kink's energy at which the state quickly disappears due to the coherence factors and the onset of strong scattering on the involved mode. We therefore estimate $\Omega_0=29\pm4$ meV for the as-grown sample, displayed in Fig. 3(C).

\subsection{Tight Binding Parameters}

The bare in-plane band structure of Bi$_2$Sr$_2$CaCu$_2$O$_{8+\delta}$ is approximated by the tight-binding formula:

$E_{A,B}(k) =\mu - 2t (\cos k_x + \cos k_y) + 4t' \cos k_x \cos k_y - 2t'' (\cos 2k_x + \cos 2k_y) \pm t_{\perp} (\cos k_x - \cos k_y)^2 /4$, 

where the index A (B) is for anti-bonding (bonding) state and $\mu$ is chemical potential. The hopping parameters that best describe the Fermi surfaces of the measured samples are kept fixed at $t=0.36$, $t'=0.108$, $t''=0.036$ and $t_{\perp}=0.108$ eV, with only the chemical potential being varied from 0.467 eV, for the non-superconducting sample to 0.425 eV, for the $T_\mathrm{c}=72$ K sample.
The TB contours that agree with the experimental contours the best were chosen by eye. By changing them to the point where discrepancies would become clearly visible, we can estimate that the uncertainty in doping, $\Delta p_\mathrm{A}$, of this method is very close to that estimated from the experimental momentum width of the Fermi surface, $\Delta p_\mathrm{A}/p_\mathrm{A}\sim2\Delta k_\mathrm{F}/k_\mathrm{F}$ .

\subsection{Other candidates for the observed renormalization}

In the following, we discuss some other possibilities for the renormalization effects observed in the antinodal region of Bi2212. One candidate with the proper trend that mimics the kink's energy is the position of van Hove singularity (vHS) of the antibonding band. A significant amount of interband scattering (elastic or inelastic) would affect the lifetime of the probed bonding state as the vHS of the antibonding state moves with doping.  However, the interband scattering would have an opposite effect of what has been seen: the interband channel (if important) would make the state broad(er) where it is open and the state would be narrower where the channel is closed (below the vHS of the antibonding band) Also, as can be seen in Fig. 2G, the kink is significantly deeper than the renormalized bottom of the antibonding band.  In addition, just as with phonons, the effect should not disappear in the normal state.

Another candidate that could possibly have similar effects on the measured quasiparticle dispersion and its lifetime is the superconducting gap itself. The observed $\Omega_0$ is very close to $\Delta_0$ and the reduction of a phase space for scattering related to the opening of the gap, would make the states sharp within a certain energy range, with details depending on the gap symmetry. In the $s$-wave gap, the kink should appear at $\sim3\Delta_0$ (or $\sim2\Delta_0$, measured from the top of quasiparticle dispersion $\Delta_0$), if it was caused by the pair-breaking. This might not be strictly valid for the $d$-wave gap, where the scattering could involve the node-antinode mixing.  However, the strength of the antinodal kink weakens rapidly as one moves from the antinode, implying that the mode scatters antinode to the antinode. Therefore, the mode's momentum has to be either $Q\approx0$, or $Q\approx(\pi,\pi)$, effectively excluding the node to antinode mixing and the pair-breaking as its origin.

\section*{Data availability} 
The data that support the findings of this study are available from the corresponding author upon reasonable request. 

\bibliographystyle{apsrev}
\bibliography{Cuprates}

\section*{Acknowledgements}
We thank I. Pletikosi\'{c}, J. Tranquada and P. Johnson for discussions. 
This work was supported by the US Department of Energy, Office of Basic Energy Sciences, contract
no. DE-AC02-98CH10886.

\section*{Author information}
\subsection{Affiliations}

Condensed Matter Physics and Materials Science Department, Brookhaven National Lab, Upton, New York 11973, USA

T. Valla, I. K. Drozdov and G. D. Gu\\

\subsection{Contributions}
T.V. designed and directed the study, performed the ARPES experiments, analyzed and interpreted data and wrote the manuscript. G.D.G. grew the bulk crystals. I.K.D. performed the sample preparation in ozone. I.K.D. and T.V. made contributions to development of the OASIS facility used herein and commented on the manuscript.

\subsection{Competing interests}
The authors declare no competing interests.

\subsection{Corresponding author}
Correspondence should be addressed to T.V. (valla@bnl.gov).

\end{document}